\documentclass[pra,showpacs,superscriptaddress,twocolumn]{revtex4}
\usepackage{dcolumn}
\usepackage{amssymb,amsmath}
\usepackage{graphicx}

\begin{document}

\raggedbottom

\title{High-resolution imaging of velocity-controlled molecular collisions using counterpropagating beams}

\author{Sjoerd N. Vogels}
\affiliation{Radboud University Nijmegen, Institute for Molecules and Materials, Heijendaalseweg 135, 6525 AJ Nijmegen, the Netherlands}

\author{Jolijn Onvlee}
\affiliation{Radboud University Nijmegen, Institute for Molecules and Materials, Heijendaalseweg 135, 6525 AJ Nijmegen, the Netherlands}

\author{Alexander von Zastrow}
\affiliation{Radboud University Nijmegen, Institute for Molecules and Materials, Heijendaalseweg 135, 6525 AJ Nijmegen, the Netherlands}

\author{Gerrit C. Groenenboom}
\affiliation{Radboud University Nijmegen, Institute for Molecules and Materials, Heijendaalseweg 135, 6525 AJ Nijmegen, the Netherlands}

\author{Ad van der Avoird}
\affiliation{Radboud University Nijmegen, Institute for Molecules and Materials, Heijendaalseweg 135, 6525 AJ Nijmegen, the Netherlands}

\author{Sebastiaan Y.T. van de Meerakker}
\affiliation{Radboud University Nijmegen, Institute for Molecules and Materials, Heijendaalseweg 135, 6525 AJ Nijmegen, the Netherlands}

\date{\today}

\begin{abstract}
We present ultrahigh-resolution measurements of state-to-state inelastic differential cross sections for NO-Ne and NO-Ar collisions, obtained by combining the Stark deceleration and velocity map imaging techniques. We show that for counterpropagating crossed beam geometries, the effect of the velocity spreads of the reagent beams on the angular resolution of the images is minimized. Futhermore, the counterpropagating geometry results in images that are symmetric with respect to the relative velocity vector. This allows for the use of inverse Abel transformation methods that enhance the resolution further. State-resolved diffraction oscillations in the differential cross sections are measured with an angular resolution approaching 0.3$^\circ$. Distinct structures observed in the cross sections gauge the quality of recent \emph{ab initio} potential energy surfaces for NO-rare gas atom collisions with unprecedented precision.
\end{abstract}

\maketitle

The development of experimental methods to study molecular collisions with the highest possible detail has been a quest in molecular collision research since it was established in the 1950s. In recent years, velocity map imaging (VMI) has revolutionized our ability to study molecular collision processes in crossed beam experiments \cite{Ashfold:PCCP8:26}. The power of the technique derives from the ability to record a two-dimensional image of scattered molecules that directly reflects their recoil velocities and thus the differential cross section (DCS) of the scattering process. In the last decade, VMI has been implemented in many scattering experiments and has made experiments possible that would have been inconceivable only a few years ago \cite{Lorenz:SCIENCE293:2063,Eyles:NatChem3:597,Lin:SCIENCE300:966,Mikosch:Science319:183}.

Precise measurements of DCSs provide unique and sensitive tests for the underlying potential energy surfaces (PESs) that govern the dynamics of interacting molecules. To validate the ever more sophisticated and accurate theoretical descriptions of scattering processes, it is essential to obtain ever higher resolutions in the experimental scattering images. The inherent resolution of VMI in principle allows for DCS measurements that probe PESs with spectroscopic resolution. The image resolution that is typically obtained in crossed beam experiments, however, is limited by the velocity spreads of both beams involved. The overlap between the signals from collision partners with different velocities can significantly blur the images, thereby obscuring detailed structures that may be present in the DCS.

In the last decade, several techniques have been developed to obtain complete control over molecules in beams. In particular, methods such as Stark and Zeeman deceleration allow for the production of packets of molecules with high density, tunable velocity, near-zero velocity spread and almost perfect quantum state purity \cite{Meerakker:CR112:4828}. Clearly, these tamed molecular beams are ideally suited for precise molecular scattering experiments, in particular when combined with state-of-the art detection techniques such as VMI. Recently, we have reported the first crossed beam experiment in which a Stark decelerator was combined with VMI, indeed resulting in scattering images with much higher resolution \cite{Zastrow:NatChem6:216,Onvlee:PCCP16:15768}.

Yet, thus far only one of the reagent beams has been velocity-controlled using a Stark decelerator. The image resolution is then limited by the relatively large velocity spread of the collision partner -- which is produced via conventional beam methods. Although reduction of this spread using mechanical velocity filters appears feasible, it will be difficult to reach the narrow distribution that is readily achieved for the Stark-decelerated reagent beam.

Here, we present a method to minimize the contribution of the collision partner's velocity spread to the angular resolution of the scattering images. Exploiting the favorable kinematics present when counterpropagating beams are used, we obtain record-high angular resolutions and thereby probe DCSs with unprecedented accuracy. This is illustrated by fully resolving diffraction oscillations in state-to-state DCSs for NO-Ne and NO-Ar collisions. For NO-Ar, we observe distinct features in the diffraction pattern that appear extremely sensitive to the details of the PESs. These features are used to gauge the quality of two recent PESs for this benchmark system.

\begin{figure}[!htb]
    \centering
    \resizebox{0.8\linewidth}{!}
    {\includegraphics{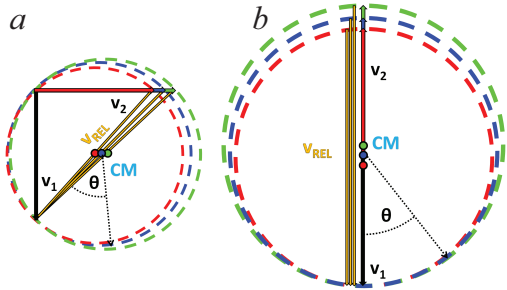}}
    \caption{Newton diagrams describing the scattering of two beams with particles of equal mass and equal mean pre-collision speed, for a beam crossing angle of 90$^{\circ}$ (\emph{a}) and 180$^{\circ}$ (\emph{b}). The beam with mean velocity $v_1$ is assumed to have zero velocity spread, whereas the other beam has a large velocity spread. A structured DCS is assumed as indicated by stripes. The relative velocity vectors ($v_{rel}$), the center-of-mass positions (CM), and the scattering deflection angle $\theta$ are indicated. We use the convention that $\theta = 0^{\circ}$ and $\theta = 180^{\circ}$ corresponds to forward and backward scattering, respectively. Note the different ring radii for the two situations, reflecting the higher collision energy for a counterpropagating geometry.}
    \label{fig:kinematics}
\end{figure}

The influence of the crossed beam geometry on the kinematics and resolution of the experiment is illustrated in Figure \ref{fig:kinematics}. For the sake of the argument, Newton diagrams are shown for the scattering of two beams containing particles of equal mass and equal pre-collision mean velocity. Beam crossing angles of 90$^\circ$ and 180$^{\circ}$ are assumed in panels \emph{a} and \emph{b}, respectively. We approximate the kinematics present in our crossed beam experiments (a Stark-decelerated beam with mean velocity $v_1$ crosses with a conventional beam with mean velocity $v_2$) by assuming that only the conventional beam has a non-zero velocity spread. To model a structured DCS, we consider a hypothetical DCS that exhibits a series of block functions, indicated by colored stripes, that are spaced with an angular interval of 12.5$^{\circ}$. Newton diagrams are shown for scattering with the mean velocity (blue), and the two outermost values from the beam distribution (red and green).

It is seen that for scattering angles around forward scattering ($\theta \approx 0^{\circ}$) -- where the most detailed structures in DCSs are typically found -- the \emph{velocity} spread of the collision partner does not contribute to the image blurring when a counterpropagating beam geometry is used. The angular image resolution is then exclusively determined by the \emph{angular} spread of the beam, and high angular image resolutions can be achieved by simply collimating the beam. This sharply contrasts the situation for a 90$^{\circ}$ crossing angle. Here, the velocity spread of the collision partner results in both radial \emph{and} angular blurring of the image; angular image resolutions can only be significantly improved by reducing the velocity spread of the collision partner.

The counterpropagating geometry has the additional advantage that scattering images are inherently cylindrically symmetric with respect to the mean relative velocity vector of the colliding beams (see Figure \ref{fig:kinematics}). This allows for the use of inverse Abel transformation methods to retrieve the angular scattering distributions that underly the three-dimensional Newton spheres. This transformation effectively cancels the additional blurring of images that occurs when these Newton spheres are projected onto the detector plane \cite{Eppink:RSI68:3477}. Again, this sharply contrasts the situation for a 90$^{\circ}$ crossing angle. Here, the inherent angular and radial resolution is different in every part of the image, and symmetry with respect to the mean relative velocity is broken. Further asymmetry in image intensity arises that is related to a detection bias for low laboratory recoil velocities which are preferentially found on one side of the image \cite{Lorenz:PCCP2:481}. Together, these asymmetries have thus far prevented the use of Abel inversion methods in crossed beam scattering experiments. Methods to improve image resolutions by recording only a central slice of the Newton sphere have been developed \cite{Townsend:RSI74:2530,Lin:RSI74:2495}, but in crossed beam experiments one cannot always afford the inherent reduction in signal intensity. Either way, complex forward iterative methods are required to disentangle instrument dependent detection probabilities from the desired scattering intensity distribution.

The use of counterpropagating beam geometries in crossed beam experiments is rather unconventional. Henning Meyer successfully used counterpropagating pulsed beam scattering techniques in the early 90's to facilitate measurements of angular resolved state specific product distributions using time-of-flight detection of the scattered molecules \cite{Meyer:JCP101:6686}, but the technique has hardly been used ever since.

We use a crossed beam scattering apparatus in which a packet of NO ($X\,^2\Pi_{1/2}, J=1/2, f$) radicals with a mean velocity of 350~m/s, a velocity spread of 1.2 m/s (1$\sigma$), and an angular spread of 0.04$^{\circ}$ (1$\sigma$) is produced by passing a beam of NO through a 2.6~m long Stark decelerator \cite{Onvlee:PCCP16:15768}. The packet of NO collides with a beam of Ne or Ar atoms at a beam intersection angle of 180$^{\circ}$, resulting in a relatively high collision energy of 720 cm$^{-1}$ and 725 cm$^{-1}$ for NO-Ne and NO-Ar, respectively. The rare gas atom beam is well collimated by placing a 1~mm wide slit between the 2~mm diameter skimmer and the interaction region. Scattered molecules are detected state-selectively using a two-color laser ionization scheme and a standard VMI geometry \cite{Eppink:RSI68:3477,Onvlee:PCCP16:15768}. The two laser beams are directed parallel to the detector plane, and cross each other at right angles.

The implementation of a counterpropagating beam geometry in our experiment is aided by the use of reagent beam pulses with a narrow temporal distribution, such that the pulses only spatially overlap within a well defined volume. In the interaction region, the packet of NO radicals and the rare gas atom beam has a temporal width of 25~$\mu$s and 15 $\mu$s, respectively. The experiment is synchronized such that the leading edges of both beams overlap for only about 5~$\mu$s before both detection lasers are fired.

\begin{figure}[htb]
    \centering
    \resizebox{0.8\linewidth}{!}
    {\includegraphics{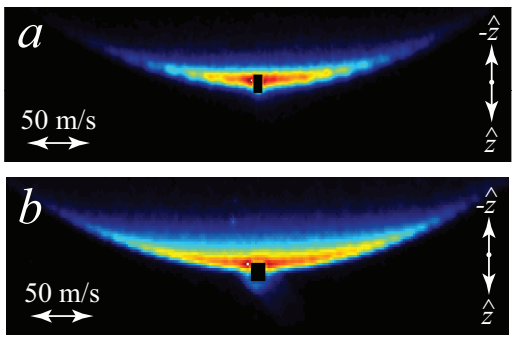}}
    \caption{Parts of experimental ion images for inelastic collisions of NO ($j=1/2,f$) with Ne (\emph{a}) and Ar (\emph{b}) atoms, exciting the NO radicals to the $j=3/2, e$ and $j=5/2,e$ state, respectively. The NO and rare gas atom beams propagate in the $\hat{z}$ and $-\hat{z}$ directions, respectively. Small segments of the distributions around forward scattering are masked due to imperfect state selection of the NO packet.}
    \label{fig:raw-data}
\end{figure}

Figure \ref{fig:raw-data}(\emph{a}) and (\emph{b}) depict the experimental scattering images for inelastic collisions of NO ($X\,^2\Pi_{1/2}, j=1/2, f$) with Ne and Ar that excite the NO radicals to the $j = 3/2,e$ and the $j=5/2,e$ state, respectively. A clear oscillatory structure -- originating from the quantum mechanical nature of molecules that leads to diffraction of matter waves during molecular collisions -- is observed for both scattering systems. The $j=3/2, e$ channel for NO-Ne has a relatively large integral cross section of 2.7 \AA$^2$, and the well-separated diffraction oscillations in this system could be observed in our previous experiments using a $90^{\circ}$ beam crossing angle \cite{Zastrow:NatChem6:216}. Comparison with the present results shows that when a counterpropagating geometry is used, a superior angular resolution is obtained, even though the collision energy is higher resulting in a smaller peak-to-peak separation of the diffraction peaks. The $j=5/2, e$ channel for NO-Ar has an integral cross section of only 0.77 \AA$^2$, and despite various efforts we have been unable to resolve oscillations in the DCS for this system in previous experiments. The channel displays one of the richest diffraction structures that we have identified thus far, with a peak-to-peak separation between individual diffraction peaks of 1.6$^{\circ}$ (\emph{vide infra}). The clearly resolved oscillations observed here for this system are a testimony of the unprecedented angular resolution afforded by the counterpropagating geometry.

As expected, the images are (almost) symmetric with respect to the relative velocity vector. Within our experimental geometry, Doppler and collision induced alignment effects cause a small dependence of the detection probability on the angle between the recoiling molecule and the propagation direction of the excitation laser. However, this asymmetry in intensity does not affect the inherent radial and angular resolution on either side of the image, and can easily be accounted for from the known geometry of the experiment. We find that the images are symmetric after correction for the Doppler effect; no indications were found that collision induced alignment plays a significant role for the inelastic channels studied here. A detection bias for low laboratory recoil velocities results in an additional image intensity correction factor that is equal on either side of the image.

\begin{figure}[!htb]
    \centering
    \resizebox{0.8\linewidth}{!}
    {\includegraphics{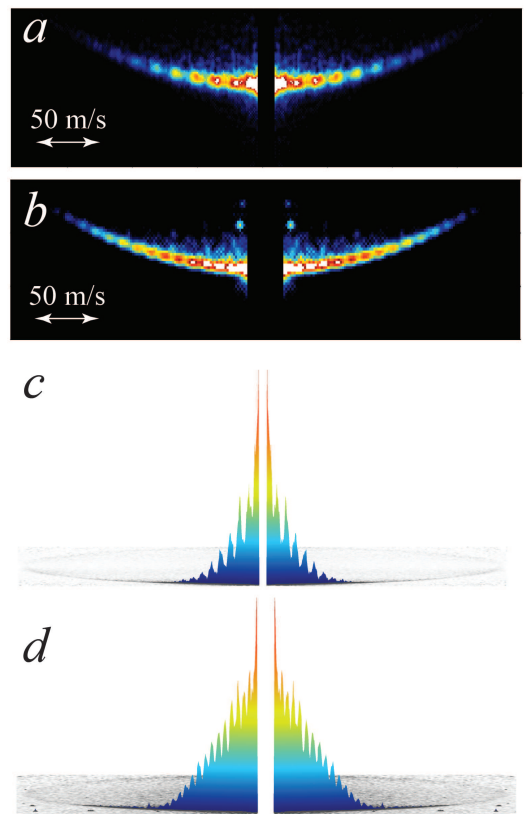}}
    \caption{Abel-inverted versions of the experimental ion images of Figure \ref{fig:raw-data}, together with the  corresponding three-dimensional representation of the scattering distributions, for NO-Ne (\emph{a} and \emph{c}) and NO-Ar (\emph{b} and \emph{d}) collisions, respectively.}
    \label{fig:Abel-data}
\end{figure}

\begin{figure}[htb]
    \centering
    \resizebox{\linewidth}{!}
    {\includegraphics{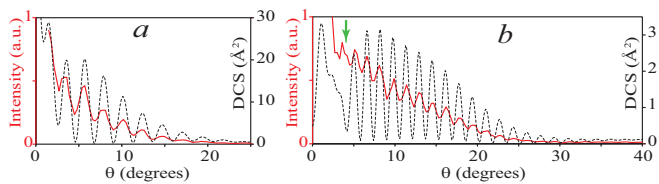}}
    \caption{Experimental angular scattering distributions (red curves) for inelastic collisions of NO ($j=1/2,f$) with Ne (\emph{a}) and Ar (\emph{b}) atoms, exciting the NO radicals to the $j=3/2, e$ and $j=5/2,e$ state, respectively, together with the theoretically predicted DCSs by Cybulski \emph{et al.} (black dashed curves).}
    \label{fig:curves}
\end{figure}

After correction, the symmetric images allow for the use of inverse Abel transformation methods to enhance the resolution further. Using the BASEX suite of programs \cite{Dibrinski:RSI73:2634}, images as shown in Figure \ref{fig:Abel-data}\emph{a} and \emph{b} are obtained for the two scattering systems. The oscillatory structure is retrieved with a much higher resolution, as is also clear from the three-dimensional representation of the data shown in panels \emph{c} and \emph{d}. Figure \ref{fig:curves} shows the angular intensity distributions (red curves) that are inferred from the experimental image intensities within a narrow annulus around the rims of the images. The DCSs at the mean collision energies of the experiment, that are predicted by high level \emph{ab initio} quantum mechanical close-coupling (QM CC) calculations that use the most recent NO-Ne and NO-Ar PESs presently available \cite{Cybulski:JPCA116:7319}, are shown as well (black dashed curves).

The experimentally obtained scattering distributions are fully consistent with the DCSs predicted by theory. In particular the angles at which diffraction peaks are found are in excellent agreement with the theoretical predictions. By definition, the theoretical DCSs presented in Figure \ref{fig:curves} predict the angular distributions that would be measured in the `perfect' experiment in which no beam spreads or other factors that limit experimental resolution are present. It is seen that in our experiments, beam spreads and other experimental factors reduce the visibility of the diffraction structure by about a factor 2 - 3 only. We estimate the angular resolution of the experiment by convoluting the DCSs predicted by the QM CC calculations with a Gaussian distribution of variable width. Best agreement with the experimental distributions is found for widths ($1 \sigma$) of 0.45$^{\circ}$ and 0.35$^{\circ}$ for NO-Ne and NO-Ar, respectively.

\begin{figure}[htb]
    \centering
    \resizebox{\linewidth}{!}
    {\includegraphics{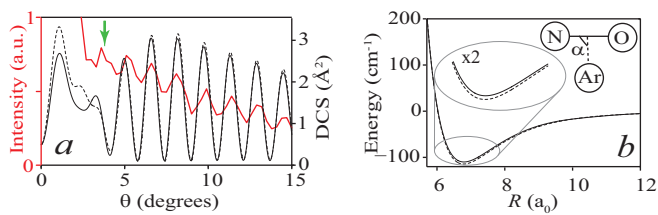}}
    \caption{\emph{(a)} Enlarged section of the experimental scattering distribution for NO-Ar collisions, together with the DCS predicted by the Cybulski (black dashed curve) and Alexander (black solid curve) PESs. \emph{(b)} Radial potential energy curves for the equilibrium geometry with $\alpha = 92^{\circ}$ for the Cybulski (dashed curve) and Alexander PESs (solid curve). Small differences in the repulsive part of the potential are present but not visible at this scale.}
    \label{fig:small-peak}
\end{figure}

The observed diffraction structure is an excellent proxy for the quality of computed PESs. For the $j = 5/2, e$ channel of NO-Ar, there is one specific peak observed at $\theta \sim 3.6^{\circ}$, as indicated by the green arrow, that appears to be extremely sensitive to the exact details of the PES. This is illustrated in Figure \ref{fig:small-peak}\emph{a} that shows the scattering intensity around forward scattering for NO-Ar on an enlarged scale. The experimentally obtained distribution is compared with the distributions that are expected from two different NO-Ar PESs: one developed by Cybulski \emph{et al.} in 2012 and one by Alexander from 1999 \cite{Alexander:JCP111:7426}. Both potentials are of high quality, but minor differences in calculation methodology and fitting procedures result in two slightly different PESs, as illustrated in panel \emph{b}. It is seen that the existence of this peak in the DCS is predicted by only one of the PESs, indicating that the level of precision reached in our experiments is sufficient to discriminate between the most state-of-the-art theoretical methods presently available. Surprisingly, our experiments show best agreement with the oldest PES used.

In this Letter, we have shown that in crossed beam scattering experiments, the ability to resolve structures in DCSs can be significantly enhanced using a beam crossing angle of 180$^{\circ}$. For scattering angles around forward scattering, the contribution of the velocity spread of the beam(s) to the angular resolution is effectively eliminated. The beneficial kinematics afforded by the counterpropagating geometry allows for the use of inverse Abel transformation methods, facilitating the measurement of DCSs with high sensitivity and angular resolution. This allows for the observation of detailed structures in the DCSs that provide unique and stringent tests to gauge the quality of PESs and quantum scattering calculations.

Our approach appears particularly beneficial in scattering experiments that employ Stark-decelerated beams, but crossed beam experiments using two conventional beams may also benefit from a 180$^{\circ}$ crossing angle. It is noted that the arguments outlined here for a counterpropagating geometry also hold for merged beam scattering, i.e., experiments that use a beam crossing angle of 0$^{\circ}$ \cite{Henson:Science338:234}. This offers prospects for achieving optimal image resolutions in experiments that aim for low collision energies; an interesting approach to probe DCSs at collision energies where the scattering is dominated by quantum scattering resonances.

\section{acknowledgements}
This work is part of the research program of the Foundation for Fundamental Research on Matter (FOM), which is financially supported by the Netherlands Organisation for Scientific Research (NWO). S.Y.T.v.d.M. acknowledges support from NWO via a VIDI and a TOP grant, and from the European Research Council via a Starting Grant. We thank Sean Gordon for assistance during a research visit and David Parker for fruitful discussions. The expert technical support by Leander Gerritsen, Chris Berkhout, Peter Claus, Niek Janssen and Andr\'e van Roij is gratefully acknowledged.


\end{document}